\documentclass[useAMS,usenatbib]{mn2e}

\usepackage{lscape}
\usepackage{tabularx}
\usepackage{graphicx}
\usepackage{amsfonts,amsmath,amssymb}
\usepackage{url}
\usepackage{enumerate}

\newcommand\ion[2]{#1{\sc #2}}%  ion, i.e. CII = \ion{C}{ii}
\newcommand\fion[2]{$[$#1{\sc #2}$]$}%  forbidden line: [CII] = \fion{C}{ii}

\title[A search for thermal X-ray signatures in GRBs II]{A search for thermal X-ray signatures in Gamma-Ray Bursts II: The \emph{Swift} sample}
\author[Sparre and Starling]{Martin Sparre$^{1,2}$\thanks{E-mail:
sparre@dark-cosmology.dk} and Rhaana L. C. Starling$^2$\\
$^{1}$Dark Cosmology Centre, Niels Bohr Institute, University of Copenhagen, Juliane Maries Vej 30, 2100 Copenhagen, Denmark\\
$^{2}$Department of Physics and Astronomy, University of Leicester, University Road, Leicester, LE1 7RH, UK}

\begin{document}

\date{\today}

\pagerange{\pageref{firstpage}--\pageref{lastpage}} \pubyear{2011}

\maketitle

\label{firstpage}

\begin{abstract}
In several gamma-ray bursts (GRBs) excess emission, in addition to the standard synchrotron afterglow spectrum, has been discovered in the early time X-ray observations. It has been proposed that this excess comes from black body emission, which may be related to the shock break-out of a supernova in the GRBs progenitor star. This hypothesis is supported by the discovery of excess emission in several GRBs with an associated supernova. Using mock spectra we show that it is only likely to detect such a component, similar to the one proposed in GRB~101219B, at low redshift and in low absorption environments. We also perform a systematic search for black body components in all the GRBs observed with the \emph{Swift} satellite and find six bursts (GRB~061021, 061110A, 081109, 090814A, 100621A and 110715A) with possible black body components. Under the assumption that their excess emission is due to a black body component we present radii, temperatures and luminosities of the emitting components. We also show that detection of black body components only is possible in a fraction of the \emph{Swift} bursts.
\end{abstract}

\begin{keywords}
gamma-rays: bursts, gamma-rays: observations, supernovae: general
\end{keywords}

\section{Introduction}

Gamma-ray bursts (GRBs) emit extreme amounts of $\gamma$-rays on a short time scale; typically
$10^{50}-10^{54}$ erg are released in $0.1-100$ seconds. Only violent processes, such as a compact object merger or the collapse of a massive star, can explain these large
energy releases, which have made GRBs observable out to high redshifts of $z\approx 8-9$ \citep{2009Natur.461.1254T,2009Natur.461.1258S,2011ApJ...736....7C}.

There exists strong evidence, that the collapse of massive stars can produce long GRBs \citep[$T_{90}>2$ s;][]{1993ApJ...413L.101K}, since
spectroscopic features from supernovae (SNe) have been detected in optical follow-up observations of GRBs (e.g. \citet{1998Natur.395..670G,2003Natur.423..847H,2011MNRAS.411.2792S}. Also see review by \citet{2011arXiv1104.2274H}). All these SNe are of type Ic with broad lines and no signs of Hydrogen
or Helium. Besides these spectroscopic detections, evidence for SN Ic features is also found in light curves of some GRBs \citep[][]{2001ApJ...555..900P,2004ApJ...606..381L,2010ApJ...718L.150C,2011MNRAS.413..669C}.

One burst of particular interest was GRB~060218 \citep{2006Natur.442.1014S,2006Natur.442.1018M}. It had an associated supernova \citep{2006Natur.442.1011P}, and its X-ray afterglow could best be described by a combination of synchrotron emission, which is usual for afterglows, and black body emission \citep{2006Natur.442.1008C}. \citet{2007ApJ...667..351W} showed that this black body emission could origin in a shock generated by the breakout of a supernova through the surface of the GRBs progenitor star. Subsequently, thermal X-ray emission which may be described by a black body
has been suggested in GRB~090618, GRB~100316D and GRB~101219B, which all have associated SNe (\citet{2011MNRAS.416.2078P}, \citet{2011MNRAS.411.2792S}, \citet{Starling2012}, respectively. See also \citet{IAU}). This supports the connection of the black body component with emission from a supernova. Deviations from a single power law in the early X-ray spectra in GRBs, was also found by \citet{2007ApJ...656.1001B}, who identified a soft emission component in $5-10$ \% of the bursts in the studied sample.

In this series of papers, we search for more bursts with X-ray black body components, and derive the conditions under which such components may be reliably recovered. In Paper I \citep{Starling2012} black body components were identified in bursts with spectroscopic or photometric signatures in the optical. The aim of this paper is to perform a systematic search for more bursts with X-ray black body components in the \emph{Swift} sample, and to derive the conditions under which such components may be reliably recovered. In Section~2 the sample and model fitting are described, and in Section~3 we create simulated spectra to set constraints on the detectability of black body components. In Section~4 bursts are selected as candidates for having black body emission, and Section~5 presents and discusses the final list of candidates. Section~6 extracts physical parameters, assuming that the excess emission is black body emission, and the fraction of GRBs with probable excess emission is examined.

For the cosmological calculations we assume a $\Lambda$CDM-universe with $h_0=0.71$,
$\Omega_\mathrm{m} = 0.27$, and $\Omega_\Lambda = 0.73$. All stated errors and error bars are 90\% confident. In the plots $n_H$ is in units of $10^{22}$ cm$^{-2}$ unless stated otherwise. We will use the words \emph{thermal components} and \emph{black body components} interchangeably.

\section{Fitting models with black body components to a sample of GRBs}

\subsection{A sample of \emph{Swift} bursts}

The sample, in which we will search for X-ray black body components, consists of the GRBs observed with the \emph{\emph{Swift}} XRT \citep{2005SSRv..120..165B}, where redshifts are determined with optical spectroscopy, or simultaneous multi-band photometry, as obtained from the GROND-instrument \citep{2011A&A...526A.153K}. Furthermore
we only select bursts, where Windowed Timing mode (WT mode) observations exist, since this assures that data were taken shortly after the trigger. The sample consists of 190 bursts with the most recent burst from August 8th 2011.

We used the publicly available\footnote{Website for XRT data:
  \url{http://www.swift.ac.uk/}} WT mode data, see \citet{2009MNRAS.397.1177E}. Note, that the
sample also includes bursts not triggered by \emph{Swift}. We make use of the time-averaged WT mode spectra, which have all been created in the same manner according to Evans et al. (2009) and using the {\it Swift} software version 3.8 and the latest calibration data.

\subsection{Spectral modelling}

In order to identify black body emission for a burst, it is necessary to understand the interplay between all contributing components in the observed spectrum: the afterglow continuum, the black body emission and the absorption of the source emission by gas in the line-of-sight within the host galaxy and the Galaxy. These properties can be parametrised by the seven parameters
in Table~\ref{table:Param}. Note, that not all parameters are free: in our sample the redshift of
each burst is known, and the Galactic column density can, in most cases, be determined to within $\pm$5\% \citep{2005A&A...440..775K}.

\begin{table*}
\caption{The parameters in the four fitting models (M1-M4). It is shown whether a parameter is free
  (if marked as \emph{free}), fixed (\emph{fixed}), not included (-) or fitted to the late time
  spectrum (\emph{late fit}).}
\label{table:Param}
\centering 
\begin{tabularx}{0.9\linewidth}{lXllll} 
\hline\hline 
 & Description & M1 & M2 & M3 & M4  \\\hline
$z$     & Redshift of the GRB & fixed&fixed&fixed&fixed  \\
$n_\text{H,gal}$ & Galactic column density& fixed&fixed&fixed&fixed\\
$n_\text{H,int}$ or $n_H$& Intrinsic column density in the GRB host galaxy&free&free&late fit&late fit\\
$\Gamma$ & Photon index, defined as $F_E\propto E^{-\Gamma}$ ($F_E$ is flux)& free&free&free&free  \\
$F_\text{PL}$  & Flux of the power law component of the afterglow in a $0.3-10$ keV band    & free&free&free&free     \\
$kT$ & Black body temperature (in the rest frame of the burst) & -&free&-&free \\
$F_\text{BB}$ & Flux of the black body emission in a $0.3-10$ keV band & -&free&-&free \\
\hline\hline
\end{tabularx}
\end{table*}

For each burst we fitted four different models to the WT spectrum. These
models are also shown in Table \ref{table:Param}. Model 1 (M1) is an absorbed power law, whereas
Model 2 (M2) also includes black body emission at the redshift of the bursts. Model 3 and 4 (M3 and M4) are similar to Model 1 and 2, with the only difference, that $n_\text{H,int}$ is fitted to the late-time Photon Counting mode spectrum, which we expect to be free from additional components and therefore a reliable measure of any absorbing column, instead of from the early-time (WT) spectrum itself. The atomic data used are Solar abundances from \citet{2000ApJ...542..914W} and cross-sections from 
\citet{1996ApJ...465..487V}.

For all the WT spectra we group the spectra in 20 counts per bin and perform a
$\chi^2$-fit. For the late time observations, used to determine $n_\text{H,int}$ in Model 3
and 4, ungrouped spectra are fitted with Cash statistics \cite{1979ApJ...228..939C}.

\subsection{Selecting candidates with the F-test}\label{CandidateSelection}

The F-test \citep[e.g.;][]{1976ApJ...208..177L} is a statistical test, which gives the probability that an improvement in the reduced $\chi^2$, with the inclusion of extra
parameters, is due to an improvement in the fitting model. To test whether a black body component is present in a spectrum, we therefore use the F-test to compare Model 1 and 2 and Model 3 and 4. \citet{2002ApJ...571..545P} showed, that the F-test is not
a stringent test, but in the following section we will explicitly show, that the F-test in many cases is sufficient to recover black body components, when they are present. In Section~\ref{FirstApproach} we will select the candidates.

\section{Recovering a black body component in mock spectra of GRB 101219B}\label{Recovering}

In Section~\ref{FirstApproach} it will be found, that one of the candidates for having a black body component is GRB~101219B with a
temperature of 0.22 keV in the Model 2 fit (see Section~\ref{FirstApproach} and Table~\ref{table_candidates_all}). This value is consistent with the value found in Paper I. With a redshift of $z=0.55$ \citep{2011GCN..11579...1D} GRB~101219B is one of the most distant GRBs with a
spectroscopically detected supernova \citep{2011ApJ...735L..24S}. Now mock spectra will be used to derive the range of column densities, redshifts and afterglow parameters for which a detection of the black body component is possible, and we will use this information to refine our candidate list.

\subsection{The role of column densities}\label{nHsimulations}

To reveal the role of the column densities, the F-test significances of the black body detections are calculated for several 101219B-like mock spectra with $n_H$-values varying from $0.0$ to
$0.8 \times 10^{22}$ cm$^{-2}$. The other parameters used to create the mock spectra are identical to the recovered value in the fit of Model 2 from the real data.

\begin{figure*}
\centering
\includegraphics[width = 0.9 \textwidth]{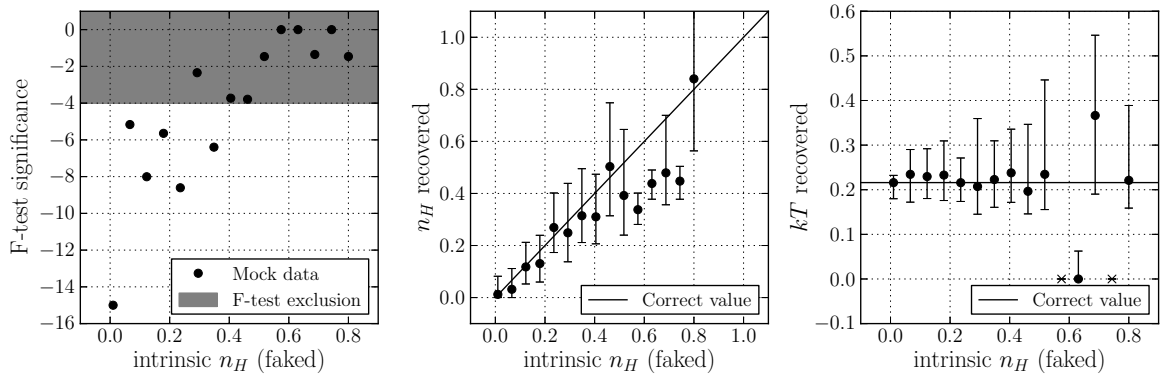}
\caption{\emph{Left}: The F-test significance of the recovery of a 101219B-like black body component as function of $n_H$ (in units of $10^{22}$ cm$^{-2}$). \emph{Central}: The $n_H$-value recovered (in the Model 2 fit) as function of the correct $n_H$. \emph{Right}: The recovered temperature (in keV).}
\label{Fig_101219B_nH}
\end{figure*}

The \emph{left panel} in Figure~\ref{Fig_101219B_nH} shows how the F-test significance (calculated from Model 1 and 2) depends on the $n_H$-value used to generate the mock spectra. If the F-test significance is lower than $10^{-4}$ we will say that Model 2 is favoured over Model 1 (i.e. the presence of a black body component is favoured). The \emph{central} and the \emph{right panel} show the $n_H$ and $kT$ recovered by the
fits of the same mock spectra. When the $n_H$ used to generate the mock data is larger than $0.4\times 10^{22}$ cm$^{-2}$, the $n_H$ recovered by the fit is often too small, and in most cases a temperature with a large error is
recovered. For $n_H<0.4 \times 10^{22}$ cm$^{-2}$ the F-test favours the existence of a black body component, and
correct $kT$ and $n_H$-values are recovered within their errors.

Similar mock spectra without black body components were also made, and no significant detections of black body components were favoured by the F-test.

\subsubsection{Fixing $n_H$ to the late time spectrum}\label{fixing}

For the fits of the real data the $n_H$-values in Model 3 and 4 are fixed to the fitted value from the late time spectrum. Now it will
explicitly be shown, that this method might lead to significant black body detections for spectra
with no real black body components, when $n_H$ is fixed to a too large value.

\begin{figure*}
\centering
\includegraphics[width = 0.95 \textwidth]{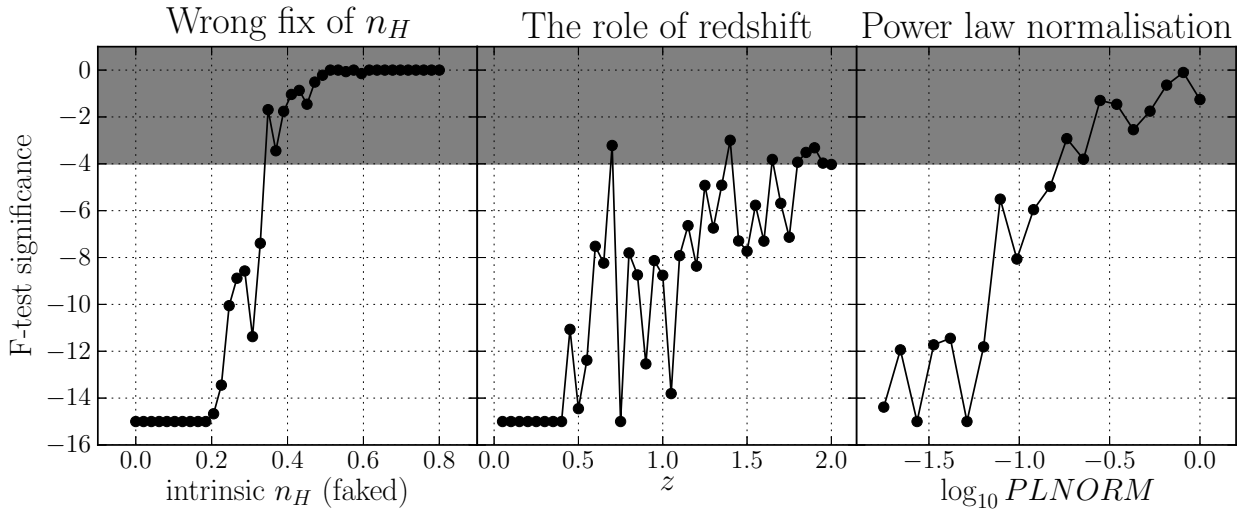}
\caption{\emph{Left}: In the fits the $n_H$-value is fixed to $0.5\times 10^{22}$ cm$^{-2}$, which is different from the $n_H$ used the generate the spectra, shown as abscissa. Spurious detections appear, when $n_H$ is fixed to a to large value. \emph{Central}: The F-test significance as function of $z$ for a GRB~101219B-like burst. \emph{Right}: The role of the power law normalization (in arbitrary units).}
\label{Fig_Wrongfix_z_exptime}
\end{figure*}

First spectra with photon indices and power law normalizations identical to GRB~101219B are generated, but
with $n_H$ varying from $0$ to $0.8\times 10^{22}$ cm$^{-2}$. All the spectra are generated without black body components. In the fitting function $n_H$ is fixed to $0.5\times 10^{22}$ cm$^{-2}$ for all the spectra, just like
if this value had come from the fit of a late time spectrum. In Figure \ref{Fig_Wrongfix_z_exptime} (\emph{left panel}) it is shown that the F-test favours a black body component when
$n_H\lesssim 0.35\times 10^{22}$ cm$^{-2}$. All the recovered black body components had temperatures between 0.08 keV and 0.2 keV. We conclude that, if $n_H$ is fixed to a value $\sim 0.15\times 10^{22}$ cm$^{-2}$ larger
than the real value, a spurious black body component will compensate for the flux lost due to the high column
density absorption parameter.

A side result of this analysis is that typical Galactic column densities can not lead to spurious detections of black body components. $n_\text{H,gal}$ is typically of order $\sim 0.05\times 10^{22}$ cm$^{-2}$ with a 10 per cent error \citep{2005A&A...440..775K}, which is well below the critical value of $ 0.15\times 10^{22}$ cm$^{-2}$ found above.

\subsection{Redshift and afterglow parameters}\label{RedshiftAndAfterglow}

To find the redshifts for which it is possible to observe a GRB~101219B-like black body component,
simulations like those in Section~\ref{nHsimulations} are performed, with the only difference
that $z$ is varied instead of $n_H$. Figure \ref{Fig_Wrongfix_z_exptime} (\emph{central panel}) shows, that black body components are recovered with F-test significances $\lesssim 10^{-8}$ when
$z<1$. For $1<z<2$ the significance of the detections are lowered by several orders of magnitudes. 

The role of the parameters describing the synchrotron emission from the afterglow will now be
quantified. First we fix the photon index to the GRB~101219B-value and simulate spectra with power
law normalizations covering the values observed in our sample. The resulting F-test significances of our fits are shown in Figure \ref{Fig_Wrongfix_z_exptime} (\emph{right panel}). A low normalization constant is
essential in order not to dilute the signal from the black body component with noise from the afterglow emission. We also fitted simulated spectra with a fixed normalisation and a varying photon index, see Figure \ref{Fig_101219B_Gamma}. A trend in the complicated pattern is that GRB~101219B-like black bodies are most likely to be detected with low photon indices.

\section{Selecting Candidates} \label{FirstApproach}

Now we will start searching for candidates with black body components in our sample. The bursts with F-test probabilities lower than $0.005$ in the Model 1-2 comparison or the Model 3-4 comparison were selected. For now, we are only interested in making an initial list of candidates, which will be refined later. We are therefore using a looser F-test significance threshold here than in Section~\ref{Recovering} (a limit of $0.005$ is now used instead of $10^{-4}$). The selected candidates are listed in Table~\ref{table_candidates_all}.

\subsection{Ruling out candidates with uncertain $n_H$}\label{unc_nh}

It is important to understand the contribution of $n_H$ in order to make a reliable detection of a black body component. We therefore removed all bursts with a 90\% error on $n_H$ larger than $0.4\times 10^{22}$ cm$^{-2}$ for Model 2 or larger than $0.8\times 10^{22}$ cm$^{-2}$ for Model 4. A lower threshold was chosen for Model 2, so candidates with a degenerate $n_H$-value were excluded. In Table~\ref{table_candidates_all} a burst is marked with \emph{uncertain $n_H$}, if it is excluded due to this criterion. The analysis was done for larger values of the thresholds, but it did not lead to the detection of more candidates in the refined analysis, which will be presented in Section~\ref{ApproachI} and \ref{ApproachII}.

\small{
		\begin{table*}

		\begin{flushleft}
\footnotetext[1]{Only for Model 2.}
			\caption{The initial candidates for having a black body component. Candidates with $z>3$ are omitted. The F-test significances are marked as $\textbf{bold}$ if the burst is a candidate in the given model. If the recovered $n_H$ is too uncertain the burst is marked with "unc. $n_H$" (see Section~\ref{unc_nh} for details).}

			\label{table_candidates_all}
			\begin{tabular}{llrrllllccccc}
				GRB&$z$&F-stat (M1-2)&F-stat (M3-4)&$kT$ (M2)&$kT$ (M4) & $n_H$ (M2)&$n_H$ (M4,fixed)&Note\\\hline
050724&0.257& $\mathbf{2.1\times 10^{-7}}$&overflow&$0.9\pm 0.1$&underflow&$0.33\pm 0.06$&$0.2 \pm 0.1$&\\
050820A&2.6147&$1.0\times 10^{-7}$&overflow&$0.4_{-0.4}^{+3.6} $&underflow&$0^{+20}_{-0}$&$0.00^{+0.09}_{-0.00}$&unc. $n_H$ High-$z$ \\
060124&2.3&$0.041$&$3.4\times 10^{-8}$&$0.6\pm 0.1$&$ 0.46\pm0.07$&$0.6 \pm 0.1$&$0.8\pm 0.2$&High-$z$ \\
060202&0.783&$\mathbf{ 4.2\times 10^{-6}}$&$\mathbf{4.3\times 10^{-14}}$&$0.38^{+0.06}_{- 0.05}$&$0.34\pm 0.03$&$1.6\pm 0.1$&$1.7\pm0.2$&\\
060218&0.0331&$\mathbf{1.0\times 10^{-174}}$&$\mathbf{9.6\times 10^{-110}}$&$0.123\pm 0.002$&$0.186\pm 0.002$&$0.76\pm0.02$&$0.40\pm0.04$&\\
060418&1.49&$\mathbf{3.0\times 10^{-6}}$&overflow&$0.61\pm 0.05$&$    0.00^{+0.03}_{-0.00}$&$0.5\pm 0.1$&$0.2^{+0.3}_{-0.2}$&\\
060502A&1.5026&$5.6\times 10^{-5}$&$\mathbf{4.7\times 10^{-5}}$&$0.35\pm 0.04$&$0.31\pm 0.02$&$0.4\pm0.2$&$0.5\pm0.2$&unc. $n_H$ (M2 only) \\
060604&2.68&$0.16$&$2.1\times 10^{-9}$&$0.15^{+0.09}_{-0.02}$&$ 1.6^{+0.3}_{-0.2}$&$2.2^{+0.7}_{-0.4}$&$1.1\pm0.3$&High-$z$ \\
060714&2.7108&$0.27$&$2.0\times 10^{-9}$&$0.24^{+0.06}_{-0.11}$&$ 1.9\pm 0.3$&$3.0 \pm 1.1$&$1.1^{+0.6}_{-0.5}$&unc. $n_H$ High-$z$ \\
060904B&0.7029&$\mathbf{0.0014}$&$    1.0$&$0.29^{+0.06}_{-0.05}$&$0.00^{+0.05}_{-0.00}$&$0.53^{+0.07}_{-0.06}$&$0.4\pm 0.1$&\\
061021&0.3463&$\mathbf{2.9\times 10^{-7}}$&$\mathbf{3.2\times 10^{-15}}$&$0.12^{+0.03}_{-0.02}$&$0.12\pm0.01$&$0.08^{+0.13}_{-0.08}$&$0.08\pm0.02$&\\
061110A&0.7578&$\mathbf{6.5\times 10^{-11}}$&$\mathbf{0.00019}$&$0.32\pm 0.02$&overflow&$0.14 \pm 0.04$&$0.3^{+0.4}_{-0.3}$&\\
061121&1.3145&$\mathbf{1.5\times 10^{-8}}$&$\mathbf{1.1\times 10^{-9}}$&$0.50 \pm 0.06$&$0.45 \pm 0.03$&$0.5\pm 0.1$&$0.61\pm 0.08$&\\
070318&0.8397&$0.0075$&$\mathbf{9.9\times 10^{-11}}$&$0.3\pm 0.1$&$0.24 \pm 0.04$&$0.6^{+0.2}_{-0.1}$&$0.8\pm 0.1$&\\
070419A&0.9705&$0.17$&$1.2\times 10^{-7}$&$ 0.4\pm 0.1$&$0.18\pm0.04$&$0.5\pm 0.1$&$0.8^{+1.7}_{-0.8}$&unc. $n_H$ \\
070508&0.82&$\mathbf{1.0\times 10^{-5}}$&$0.074$&$0.066^{+0.002}_{-0.004}$&$0.030^{+0.002}_{-0.001}$&$1.12^{+0.09}_{-0.07}$&$0.6\pm0.2$&\\
070724A&0.457&$0.15$&$\mathbf{0.0014}$&$0.15^{+0.10}_{-0.04}$&$0.8^{+0.2}_{-0.1}$&$1.0\pm 0.5$&$0.1^{+0.3}_{-0.1}$&\\
071031&2.6918&$6.1\times 10^{-9}$&$8.4\times 10^{-67}$&$ 2.1^{+0.4}_{-0.3}$&$ 1.44\pm 0.06$&$0.9 \pm 0.1$&$0.0^{+0.7}_{-0.0}$&High-$z$ \\
071112C&0.8227&$\mathbf{0.00011}$&$\mathbf{0.00013}$&$0.40^{+0.06}_{-0.07}$&$ 0.37^{+0.03}_{-0.04}$&$0.08^{+0.08}_{-0.07}$&$0.1^{+0.2}_{-0.1}$&\\
080210&2.6419&$0.10$&$0.0020$&$ 1.0^{+  0.3}_{-0.2}$&$ 1.0^{+0.3}_{-0.2}$&$1.0^{+1.2}_{-1.0}$&$1.5^{+0.8}_{-0.7}$&unc. $n_H$ High-$z$ \\
080310&2.4274&$0.052$&$1.7\times 10^{-22}$&$ 2.0^{+ 1.1}_{- 0.7}$&$ 1.3\pm 0.1$&$0.63^{+0.07}_{-0.09}$&$0.4^{+0.3}_{-0.2}$&High-$z$ \\
080319B&0.9382&$\mathbf{1.4\times 10^{-25}}$&$    1.0$&$ 1.02^{+0.08}_{-0.07}$&underflow&$0.12\pm 0.01$&$0.08^{+0.04}_{-0.03}$&\\
080413A&2.433&$0.27$&$2.0\times 10^{-7}$&$0.14^{+0.04}_{-0.03}$&$1.0\pm 0.1 $&$2.8^{+0.9}_{-0.6}$&$0.5^{+1.0}_{-0.5}$&unc. $n_H$ High-$z$ \\
080430&0.767&$\mathbf{0.00020}$&$\mathbf{3.3\times 10^{-5}}$&$0.24\pm 0.06$&$0.21\pm 0.03$&$0.3^{+0.3}_{-0.1}$&$0.36\pm 0.06$&\\
080603B&2.6892&$0.042$&$2.8\times 10^{-13}$&$ 1.7^{+0.6}_{-0.3}$&$ 1.6^{+0.2}_{-0.1}$&$0.8\pm 0.3$&$0.2^{+0.6}_{-0.2}$&unc. $n_H$ High-$z$ \\
080604&1.4171&$\mathbf{0.0010}$&$\mathbf{3.0\times 10^{-6}}$&$ 0.34^{+0.06}_{-0.10}$&$0.28\pm0.04$&$0.04^{+0.21}_{-0.04}$&$0.2^{+0.5}_{-0.2}$&\\
080605&1.6403&$\mathbf{0.00011}$&$0.84$&$0.087^{+0.020}_{-0.006}$&$0.04^{+0.09}_{-0.04}$&$1.2^{+0.2}_{-0.1}$&$0.6\pm0.3$&\\
080721&2.5914&$2.6\times 10^{-10}$&$5.3\times 10^{-51}$&$0.21^{+0.01}_{-0.02}$&$ 2.0 \pm 0.1$&$2.2 \pm 0.3$&$0.5\pm 0.2$&unc. $n_H$ High-$z$ \\
080805&1.5042&$0.38$&$1.1\times 10^{-7}$&$0.06^{+0.05}_{-0.02}$&$0.27\pm0.05$&$0.65^{+0.16}_{-0.07}$&$1.2\pm 0.5$&unc. $n_H$ \\
080928&1.6919&$\mathbf{2.9\times 10^{-5}}$&overflow&$ 1.7^{+0.4}_{-0.3}$&underflow&$0.5\pm 0.1$&$0.3\pm 0.1$&\\
081007&0.5295&$0.0016$&$\mathbf{0.00032}$&$0.20^{+0.07}_{-0.05}$&$0.30\pm 0.03$&$0.8^{+0.3}_{-0.2}$&$0.5 \pm 0.1$&unc. $n_H$ (M2 only)\\
081008&1.967&$0.057$&$\mathbf{6.7\times 10^{-33}}$&$ 2.1 \pm 0.8$&$ 1.19 ^{+0.08}_{-0.07}$&$0.9\pm 0.1$&$0.3 \pm 0.3$& \\
081109&0.9787&$0.047$&$\mathbf{1.0\times 10^{-5}}$&$1.0^{+0.3}_{-0.2}$&$0.13\pm 0.04$&$0.5\pm 0.1$&$1.0\pm 0.2$&\\
081203A&2.05&$0.00073$&$0.053$&$0.5\pm 0.08$&overflow&$1.4\pm 0.5$&$2^{+2}_{-1}$&unc. $n_H$ High-$z$ \\
081222&2.77&$0.086$&$2.9\times 10^{-5}$&$0.16^{+0.05}_{-0.07}$&$ 1.3^{+ 0.3}_{-0.2}$&$1.0^{+0.4}_{-0.3}$&$0.4\pm 0.2$&High-$z$ \\
081230&2.03&$0.12$&$0.00023$&$ 2.8^{+0.7}_{-1.0}$&$ 1.5\pm 0.4$&$0.8^{+0.4}_{-0.3}$&$0.2 \pm 0.2$&High-$z$ \\
090418A&1.608&$0.062$&$\mathbf{8.3\times 10^{-5}}$&$ 0.5\pm 0.2$&$0.27 \pm 0.09$&$0.5^{+0.6}_{-0.3}$&$1.1\pm0.2$&\\
090424&0.544&$\mathbf{0.0041}$&$\mathbf{0.0042}$&$ 1.5^{+ 0.2}_{-0.3}$&$ 1.3\pm 0.2$&$0.44^{+0.03}_{-0.02}$&$0.42\pm 0.06$&\\
090618&0.54&$\mathbf{0.00026}$&$\mathbf{4.1\times 10^{-6}}$&$ 1.9\pm 0.2$&$ 0.38^{+0.04}_{-0.03}$&$0.26\pm 0.02$&$0.19\pm 0.02$&\\
090809&2.737&$0.020$&$4.3\times 10^{-7}$&$0.21\pm 0.03$&$ 2.1\pm 0.3$&$6\pm 1$&$0.6\pm 0.3$&High-$z$ \\
090812&2.452&$1.5\times 10^{-5}$&$1.9\times 10^{-8}$&$ 1.5^{+0.5}_{-0.2}$&$0.10\pm 0.02$&$0.8^{+0.2}_{-0.1}$&$1.2^{+0.7}_{-0.6}$&High-$z$ \\
090814A&0.696&$\mathbf{1.6\times 10^{-10}}$&$\mathbf{1.3\times 10^{-10}}$&$0.28 \pm 0.03$&$0.26\pm 0.01$&$0.14 \pm 0.06$&$0.2^{+0.3}_{-0.2}$&\\
090926B&1.24&$2.9\times 10^{-9}$&$1.8\times 10^{-9}$&$0.53\pm 0.07$&$0.52 \pm 0.04 $&$1.5^{+0.3}_{-0.2}$&$1.5^{+1.2}_{-0.8}$&unc. $n_H$ \\
091029&2.752&$2.0\times 10^{-5}$&$0.0054$&$0.38^{+0.08}_{-0.07}$&overflow&$1.0^{+0.7}_{-0.5}$&$0.5\pm 0.2$&unc. $n_H$ High-$z$ \\
100117A&0.92&$0.0055$&$1.7\times 10^{-10}$&$ 1.2^{+0.4}_{-0.2}$&$0.09\pm 0.2$&$0.2 \pm 0.2$&$2.2^{+1.6}_{-1.2}$&unc. $n_H$ \\
100418A&0.6235&$\mathbf{0.00074}$&$\mathbf{2.3\times 10^{-11}}$&$0.21^{+0.03}_{-0.04}$&$0.24\pm 0.01$&$0.2\pm 0.1$&$0.1^{+0.4}_{-0.1}$&\\
100621A&0.542&$\mathbf{0.00011}$&$\mathbf{2.7\times 10^{-5}}$&$0.48^{+0.05}_{-0.06}$&$0.29^{+0.03}_{-0.04}$&$1.4 \pm 0.2$&$2.0 \pm 0.2$&\\
101219B&0.55&$\mathbf{2.2\times 10^{-13}}$&$\mathbf{3.9\times 10^{-27}}$&$ 0.22\pm 0.03$&$0.20\pm 0.1$&$0.06^{+0.06}_{-0.05}$&$0.1\pm 0.1$&\\
110205A&2.22&$0.021$&$4.7\times 10^{-10}$&$ 3 \pm 1$&$  1.6^{+0.3}_{-0.2}$&$0.54^{+0.08}_{-0.10}$&$0.4\pm 0.1$&High-$z$ \\
110715A&0.82&$0.40$&$\mathbf{6.6\times 10^{-6}}$&$ 0.15^{+0.06}_{-0.08}$&$0.15^{+0.04}_{-0.03}$&$1.0\pm 0.4$&$1.3 \pm 0.3$&\\
110731A&2.83&$0.22$&$0.0049$&$0.12^{+0.07}_{-0.04}$&$ 1.7^{+ 0.6}_{-  0.5}$&$1.8\pm 0.3$&$1.1^{+0.7}_{-0.6}$&unc. $n_H$ High-$z$ \\
110808A&1.348&$0.0017$&$0.00068$&$0.29 \pm 0.07$&$0.22^{+0.02}_{-0.03}$&$0.2^{+0.3}_{-0.2}$&$0.6^{+0.4}_{-0.3}$&unc. $n_H$ \\
			\end{tabular}
		\end{flushleft}
		\end{table*}
}

\subsection{Candidates at high and low redshift}

\subsubsection{The high-redshift candidates ($z>2$)}

For $z>2$ GRB~101219B-like black body components are difficult to detect (see Section~\ref{RedshiftAndAfterglow} and Figure~\ref{Fig_Wrongfix_z_exptime}). A number of candidates with $z>2$ have, however, emerged in our analysis. They typically exhibit high temperatures ($kT\gtrsim 1$ keV), which is much higher than most previously claimed examples \cite[e.g.;][]{2006Natur.442.1008C,2011MNRAS.411.2792S}. They also have an uncertain $n_H$, which increases the risk of a spurious detection.

At low redshift (e.g. $z<1$), only a few candidates have temperatures larger than $1$ keV, even though a detection of such a black body component is expected to be easier at lower redshift. The high-redshift candidates are therefore unlikely to represent real black bodies, and they will not be examined further in this work.

\begin{figure}
\centering
\includegraphics[width = 0.45 \textwidth]{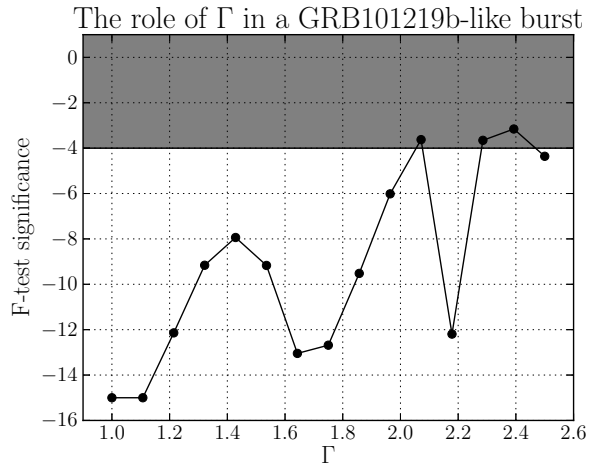}
\caption{The F-test significance as function of the photon index.}
\label{Fig_101219B_Gamma}
\end{figure}

\subsubsection{The low-redshift candidates ($z<2$)}

Three of the candidates, GRB~060218, GRB~090618 and GRB~101219B, already have identified black body components in other studies, and they also have a spectroscopically confirmed SN in the optical. Among our candidates is also GRB~100418A, which also has a SN in the optical (de Ugarte Postigo et al. in prep.). Paper I studied the early X-ray spectrum of GRB\,100418, and also find that a model with a powerlaw and a black body model can give a good fit. However, they cannot rule out an absorbed power law model given the large uncertainty on the intrinsic absorbing column of this GRB host, which may be higher than the limit for detectability we find in Section~\ref{nHsimulations} of $0.4 \times 10^{22}$ cm$^{-2}$. Paper I also studied GRB~081007, and found an indication of a black body component, but no conclusive evidence.

At first sight it is worrying, that GRB~100316D, which \citet{2011MNRAS.411.2792S} showed had an excess black body emission, is not among our candidates. An examination of this burst shows, that it is not selected due to a F-test value of 0.007 (for the Model 1-2 comparison), which is slightly above our threshold of 0.005 defined in Section~\ref{CandidateSelection}. It is therefore clear that our candidate selection algorithm will not find all bursts with black body components (discussed further in Section~\ref{fractionofbursts}).

\begin{figure}
\centering
\includegraphics[width = 0.45 \textwidth]{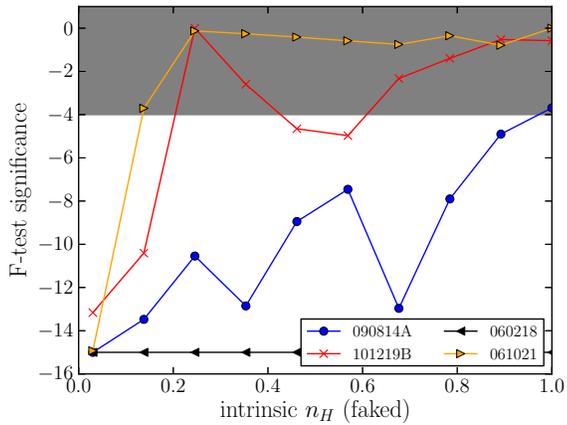}
\caption{Same as Figure~\ref{Fig_101219B_nH} (left), but with three additional bursts.}
\label{Fig_Gen1}
\end{figure}

\begin{figure}
\centering
\includegraphics[width = 0.45 \textwidth]{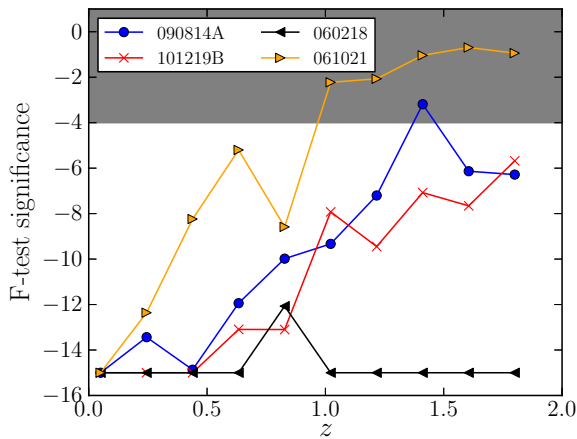}
\caption{Same as Figure~\ref{Fig_Wrongfix_z_exptime} (central), but with three additional bursts.}
\label{Fig_Gen2}
\end{figure}

Figure~\ref{Fig_Gen1} and \ref{Fig_Gen2} show how the F-test significances of the black body component detections in a selection of our low redshift candidates depend on column density and redshift. In general the same trends found for
GRB~101219B above applies for both GRB~090814A and GRB~061021. A black body component, similar to the one present in GRB~060218, is expected to be detectable at high redshifts (e.g. $z\gtrsim$ 2).

\subsection{Approach I: Fitting with 5 free parameters}\label{ApproachI}

The final step in our selection procedure is to do a time sliced analysis of the remaining candidates, i.e. those without remarks in Table~\ref{table_candidates_all}. In this section we will focus on the candidates selected in the comparison between Model 1 and 2. In Section~\ref{ApproachII} we will fit and analyse the candidates selected with Model 3 and 4.

For each of the candidates, where Model 2 is favoured by the F-test, we manually inspected the light curves, and selected time intervals, where the light curves exhibit a single power law decay. We especially tried to avoid time intervals with flares and plateaus.

To each time sliced spectrum all the models were fit, see a summary of the fits in Table~\ref{timeslice}. Several candidates with no flare-free epochs are excluded from the table. The candidates with excess emission are GRB~061021, GRB~061110A and GRB~090814A. They have low column densities ($n_H<0.3\times 10^{22}$ cm$^{-2}$) for both Model 2, where the $n_H$ is a free parameter, and Model 4, where $n_H$ is fitted with the late time spectrum. In all the fits the presence of a black body component is favoured by the F-test with significances from $0.0005$ to $2.4\times 10^{-13}$. The temperatures are between $0.1$ and $0.3$ keV, which is consistent with claimed black body components in other studies.

In Table~\ref{timeslice}, some bursts are also listed as candidates with some caveats. These are GRB~100621A (because it has a F-test significance of 0.00144, and a large $n_H$), GRB~061121 (it has large $kT$ and $z$, and the F-test does not favour a black body component in Model 3 and 4) and GRB~050724 (it has an unexpectedly high $kT$). We note that GRB~050724 has been classified as a short burst with extended emission \citep{2005Natur.438..994B}, so its origins may not lie in the collapse of a massive star as is the case for the rest of our sample which are all long GRBs.

\subsection{Approach II: Breaking $N_H$-degeneracies with the late time spectrum}\label{ApproachII}

The three bursts, which were candidates for having black body components in the previous subsection, all have low column densities $n_H<0.3\times 10^{22}$ cm$^{-2}$. This is expected from Figure~\ref{Fig_Gen1} which showed, that it is hard to recover a black body component at large $n_H$, when it is fitted as a free parameter (as in Model 1 and 2).

We will now do a new search for black body components, where we use the the late time spectrum to fit $n_H$, i.e. use Model 3 and 4. This method can break the degeneracies between $n_H$ and the other parameters, but it also has the disadvantage that a spurious black body component can be found if a too high value of $n_H$ is recovered in the late time spectrum (see Section~\ref{fixing}).

We fitted Model 3 and 4 to the candidates. Table~\ref{timeslice2} summarizes the fits. Several of the candidates are not shown in the table, since they have complex and bumpy light curves, or too little data to extract a good spectrum.

\begin{figure}
\centering
\includegraphics[width = 0.45 \textwidth]{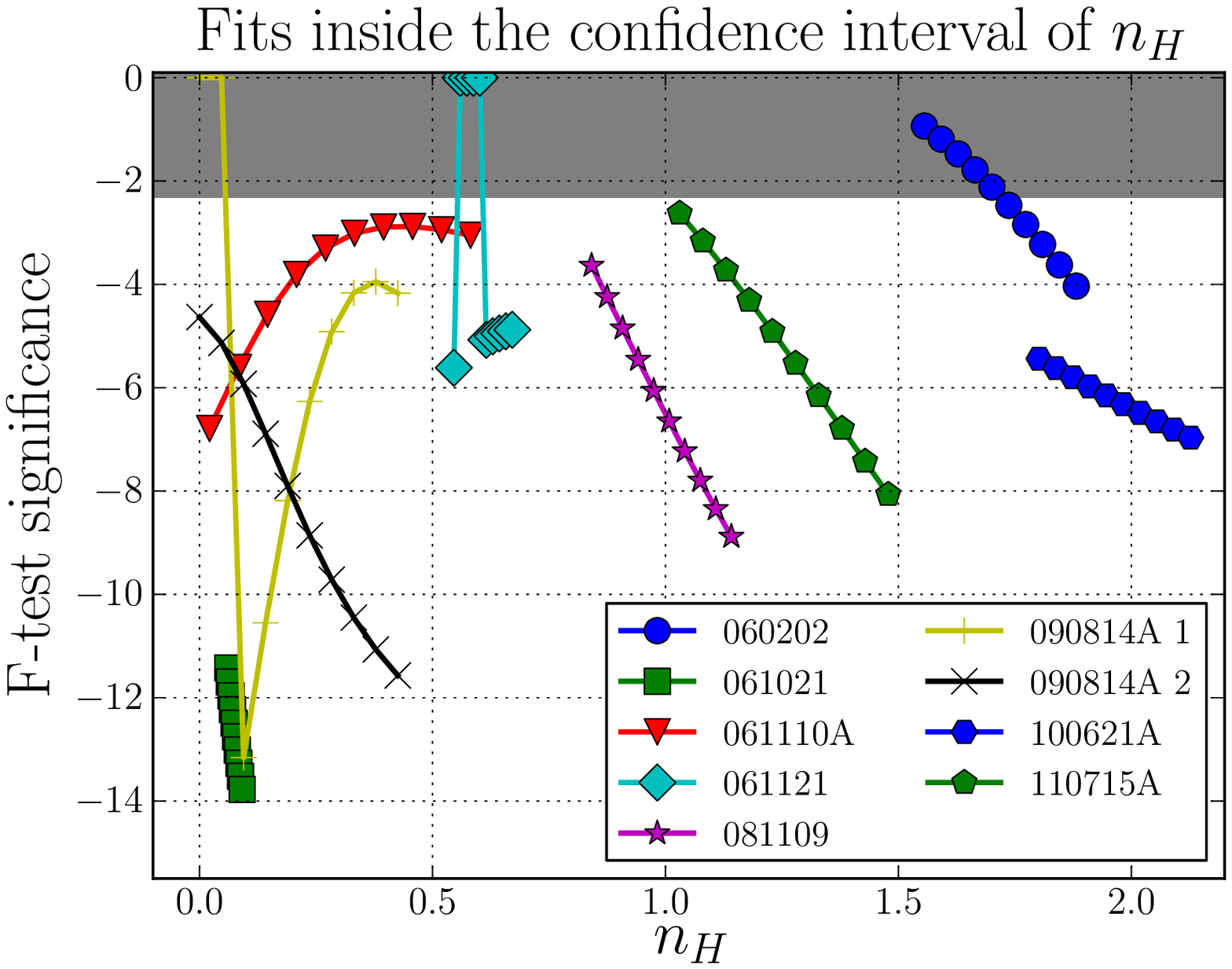}
\caption{In Model 3 and 4 $n_H$ is fixed to a value determined from late time observations. This figure examines how the F-test significances of the candidates from Table~\ref{timeslice2} depend on $n_H$-values with the $90$\% confidence intervals found in the fits of the late time spectra.}
\label{Fig_nHXConfFixed}
\end{figure}

Several bursts have favourable F-test significances. To explicitly check whether the fixing of $n_H$ might have caused spurious black body detections, we made additional fits, where $n_H$ was fixed to a range of different values within the $90\%$ confidence intervals recovered in the late time fits. Figure~\ref{Fig_nHXConfFixed} shows how the significances of the detections depend on the value to which $n_H$ is fixed. The detection is not robust for GRB~060202 and GRB~061121 because of degeneracies between $kT$ and $n_H$ from the late time spectrum. The list of candidates from the Model 3 to 4 comparison consists of GRB~061021, 061110A, 081109, 090814A, 100621A and 110715A.

		\begin{table*}

	        \begin{flushleft}		
			\caption{The time sliced fits from Section~\ref{ApproachI}, where bursts are selected with Model 1 and 2. Three reliably detected black body components are found.}\label{timeslice}

			\begin{tabular}{l}{\bf Candidates with probable excess emission:}\\\end{tabular}
			\begin{tabular}{llcccccccccc} GRB&$z$&Time&F-test M1$-$2&F-test M3$-$4&$kT$ (M2)&$kT$ (M4)&$n_H$ (M2)&$n_H$ (M4,fixed)&
			\\\hline
			061021&$0.3663$&$87-170$&$3.1\times 10^{-7}$&$2.4\times 10^{-13}$&$0.13^{+0.03}_{-0.02}$&$ 0.13\pm 0.01 $&$0.08^{+0.13}_{-0.08}$ &$0.08\pm 0.02$&
			\\
			061110A&$0.7578$&$180-240$&$2.5\times 10^{-6}$&$0.00050$&$0.25\pm 0.01$&$0.19 \pm 0.01$&$0.00^{+0.07}_{-0.00}$ &$0.3^{+0.3}_{-0.2}$
			\\
            090814A&$0.696$&$166-265$&$4.0\times 10^{-7}$&$2.2\times 10^{-9}$&$0.32\pm 0.03$&$0.30\pm 0.01$  &$0.12\pm 0.06$&$0.2^{+0.3}_{-0.2}$
            \\
            090814A&$0.696$&$265-390$&$1.1\times 10^{-6}$&$2.1\times 10^{-8}$&$0.19^{+0.05}_{-0.04}$&$0.20\pm 0.02$& $0.2^{+0.2}_{-0.1}$ &$0.2^{+0.3}_{-0.2}$&
            \\
			\end{tabular}
			\begin{tabular}{l}{\bf Candidates with some caveats:}\\\end{tabular}
			\begin{tabular}{llcccccccccc} GRB&$z$&Time&F-test M1$-$2&F-test M3$-$4&$kT$ (M2)&$kT$ (M4)&$n_H$ (M2)&$n_H$ (M4,fixed)&\\\hline
			050724&$0.257$&$100-190$&$1.3\times 10^{-6}$&$-$&$0.9\pm 0.1$&$-$&$0.36 \pm 0.09$&$0.2\pm 0.1$&\\
			061121&$1.3145$&$125-215$&$1.1\times 10^{-5}$&$1$&$0.44^{+0.05}_{-0.06}$&$0.0^{+0.1}_{-0.0}$&$0.6\pm 0.1$&$0.61 \pm 0.07 $&\\
			100621A&$0.542$&$80-120$&$0.022$&$0.86$&$0.69^{+0.06}_{-0.07}$&$0.05\pm 0.05$&$1.4^{+0.6}_{-0.3}$&$2.0\pm 0.2$&\\
			100621A&$0.542$&$190-230$&$0.0014$&$6.0\times 10^{-7}$&$0.33\pm 0.06$&$0.25\pm 0.03$&$1.3^{+0.4}_{-0.3}$&$2.0\pm 0.2$&\\
			\end{tabular}
			\begin{tabular}{l}{\bf Candidates where a black body component is not favoured:}\\\end{tabular}
\begin{tabular}{llcccccccccc} GRB&$z$&Time&F-test M1$-$2&F-test M3$-$4&$kT$ (M2)&$kT$ (M4)&$n_H$ (M2)&$n_H$ (M4,fixed)\\\hline
060202&0.783&$148-303$&$0.0023$&$0.12$&$ 0.42^{+0.07}_{-0.06}$&$0.045^{+0.003}_{-0.002}$&$1.6\pm 0.1$&$1.7\pm 0.2$&\\
060202&0.783&$745-1000$&$0.28$&$0.0058$&overflow&$ 0.25^{+0.08}_{-0.06}$&$1.6\pm 0.2$&$1.7\pm 0.2$&\\
060418&1.49&$240-450$&$ 0.29$&$  0.20$&$0.6\pm 0.2$&$0.6^{+0.2}_{-0.1}$&$0.2\pm 0.2$&$0.2^{+0.3}_{-0.2}$&\\
070508&0.82&$200-500$&$0.19$&$0.17$&$0.06^{+0.09}_{- 0.02}$&$0.031\pm 0.002$&$1.0^{+0.3}_{-0.1}$&$0.6\pm 0.2$&\\
070508&0.82&$500-1000$&$0.055$&$0.33$&$ 1.0^{+0.8}_{-0.2}$&$0.030^{+0.014}_{-0.004}$&$0.8\pm 0.2 $&$0.6 \pm 0.2 $&\\
071112C&0.8227&$89-179$&$0.022$&undef&$0.40^{+0.08}_{-0.09}$&$0.00^{+0.04}_{-0.00}$&$0.2\pm 0.1$&$0.1^{+0.2}_{-0.1}$&\\
080319B&0.9382&$200-500$&$0.0077$&$    1.0$&$1.0^{+0.4}_{-0.2}$&$0.00^{+0.03}_{-0.00}$&$0.13\pm 0.03$&$0.08\pm 0.03$&\\
080430&-\\
080604&1.4171&$125-200$&$0.14$&$0.059$&$0.21^{+ 0.15}_{-0.06}$&$0.27^{+0.06}_{-0.07}$&$0.3^{+0.4}_{-0.3}$&$0.2^{+0.5}_{-0.2}$&\\
080605&1.6403&$300-700$&$0.054$&$    1.0$&$0.084^{+0.004}_{-0.008}$&$0.003^{+0.094}_{-0.003}$&$1.3^{+0.2}_{-0.1}$&$0.6\pm 0.2$&\\
080928&1.6919&$280-320$&$0.0014$&undef&$ 2.3^{+0.9}_{-0.6}$&$0.0^{+0.1}_{-0.0}$&$0.7\pm 0.3$&$0.3\pm 0.1$&\\
090424&0.544&$700-1000$&$0.62$&$0.017$&$ 1.1^{+0.8}_{-0.4}$&$0.8\pm 0.2$&$0.52\pm 0.07 $&$0.42\pm 0.05$&\\
\end{tabular}

\end{flushleft}
\end{table*}

		\begin{table*}$\phantom{.}$
		\label{final_table}

	        \begin{flushleft}		
			\caption{The time sliced fits from Section~\ref{ApproachII}, where bursts are selected with Model 3 and 4. Three reliably detected black body components are found.}\label{timeslice2}

			\begin{tabular}{l}{\bf Candidates with probable excess emission:}\\\end{tabular}
			\begin{tabular}{llcccccccc} GRB&$z$&Time&F-test M3$-$4&$kT$ (M4)&$n_H$ (M4,fixed)&
			\\\hline
061021&0.3463&$87-170$&$2.4\times 10^{-13}$&$ 0.13\pm 0.01 $&$0.08\pm 0.02$&
\\
061110A&0.7578&$180-240$&$0.00050$&$0.19\pm 0.01 $&$0.3^{+0.3}_{-0.2}$&
\\
081109&0.9787&$90-200$&$6.3\times 10^{-7}$&$0.16\pm 0.03 $&$1.0^{+0.2}_{-0.1}$&
\\
090814A&0.696&$166-265$&$2.2\times 10^{-9}$&$  0.30\pm 0.01$&$0.3\pm 0.2$&
\\
090814A&0.696&$265-390$&$2.1\times 10^{-8}$&$0.20 \pm 0.02$&$0.2\pm 0.2$&
\\
100621A&0.542&$80-120$&$0.86$&$0.05\pm0.05$&$2.0\pm 0.2$&
\\
100621A&0.542&$190-230$&$6.0\times 10^{-7}$&$0.25\pm 0.03 $&$2.0\pm 0.2$&
\\
110715A&0.82&$97-501$&$6.6\times 10^{-6}$&$0.15\pm 0.03$&$1.3\pm 0.2$&
\\
			\end{tabular}

			\begin{tabular}{l}{\bf Candidates with some caveats:}\\\end{tabular}\\
			\begin{tabular}{llccccccc} GRB&$z$&Time&F-test M3$-$4&$kT$ (M4)&$n_H$ (M4,fixed)\\\hline
060202&0.783&$148-303$&$0.12$&$0.045^{+0.004}_{-0.002}$&$1.7\pm 0.2$&\\
060202&0.783&$745-1000$&$0.0058$&$ 0.25^{+0.07}_{-0.06}$&$1.7\pm 0.2$&\\
061121&1.3145&$125-215$&$7.5\times 10^{-6}$&$0.43\pm 0.03 $&$0.61^{+0.07}_{-0.06}$&\\
			\end{tabular}

			\begin{tabular}{l}{\bf Candidates where a black body component is not favoured:}\\\end{tabular}
			\begin{tabular}{llccccccc} GRB&$z$&Time&F-test M3$-$4&$kT$ (M4)&$n_H$ (M4,fixed)\\\hline
071112C&0.8227&$89-179$&undef&$0.00^{+0.04}_{-0.00}$&$0.1^{+0.2}_{-0.1}$&\\
080604&1.4171&$125-200$&$0.059$&$0.3\pm 0.1$&$0.2^{+0.5}_{-0.2}$&\\
090424&0.544&$700-1000$&$0.017$&$0.8\pm0.2$&$0.42\pm 0.05$&\\
			\end{tabular}

\end{flushleft}
\end{table*}

\section{The final candidates and potential caveats}\label{sec5}

We will refer to the bursts in Table~\ref{timeslice2} as our final list of candidates for having a black body component. In this section we will discuss potential caveats for each candidate.

\subsection{The Redshifts}

It is essential that the redshift, which is a fixed parameter in all our models, is reliably determined. Table~\ref{table:redshifts} summarizes how the redshift has been found for the bursts with excess emission. GRB~061021, GRB~090814A and GRB~110715 have redshifs found with only two absorption lines. It is however possible that these absorption lines comes from other galaxies than the host galaxy of the bursts. The remaining bursts have more reliably determined redshifts with several absorption and/or emission lines.

\begin{table*}

\label{table:redshifts}
\centering 
\caption{A schematic summary of how the redshift has been determined for each of the final candidates for having black body components.}
\begin{tabularx}{0.9\linewidth}{llllX} 
\hline\hline 
GRB & Redshift &Absorption lines&Emission lines&References\\
\hline
061021 & 0.3463 & \ion{Mg}{ii}2796, \ion{Mg}{ii}2803 & - & \citet{2009ApJS..185..526F}. \\
061110A & 0.7578 & \ion{Mg}{ii}2796, \ion{Mg}{ii}2803 & H$\beta$, \fion{O}{iii} doublet &\citet{2009ApJS..185..526F}.\\
081109 & 0.9787& - &GROND photometry&\citet{2011AA...534A.108K}.\\%\\
090814A&$ $$0.696$&  \ion{Mg}{ii} and \ion{Ca}{ii} &- &\citet{2009GCN..9797....1J}.\\
100621 & 0.542 &-&\fion{O}{ii} 3727, H$\beta$,
\fion{O}{iii} doublet& \citet{2010GCN..10876...1M,2011AA...534A.108K}.\\
110715A & 0.82 & \ion{Ca}{ii}, \ion{Ca}{i}&-&\citet{2011GCN..12164...1P}.\\
\hline\hline

\end{tabularx}

\end{table*}

\subsection{Light curves and spectral evolution}

\begin{figure*}
\centering
\includegraphics[width =  \textwidth]{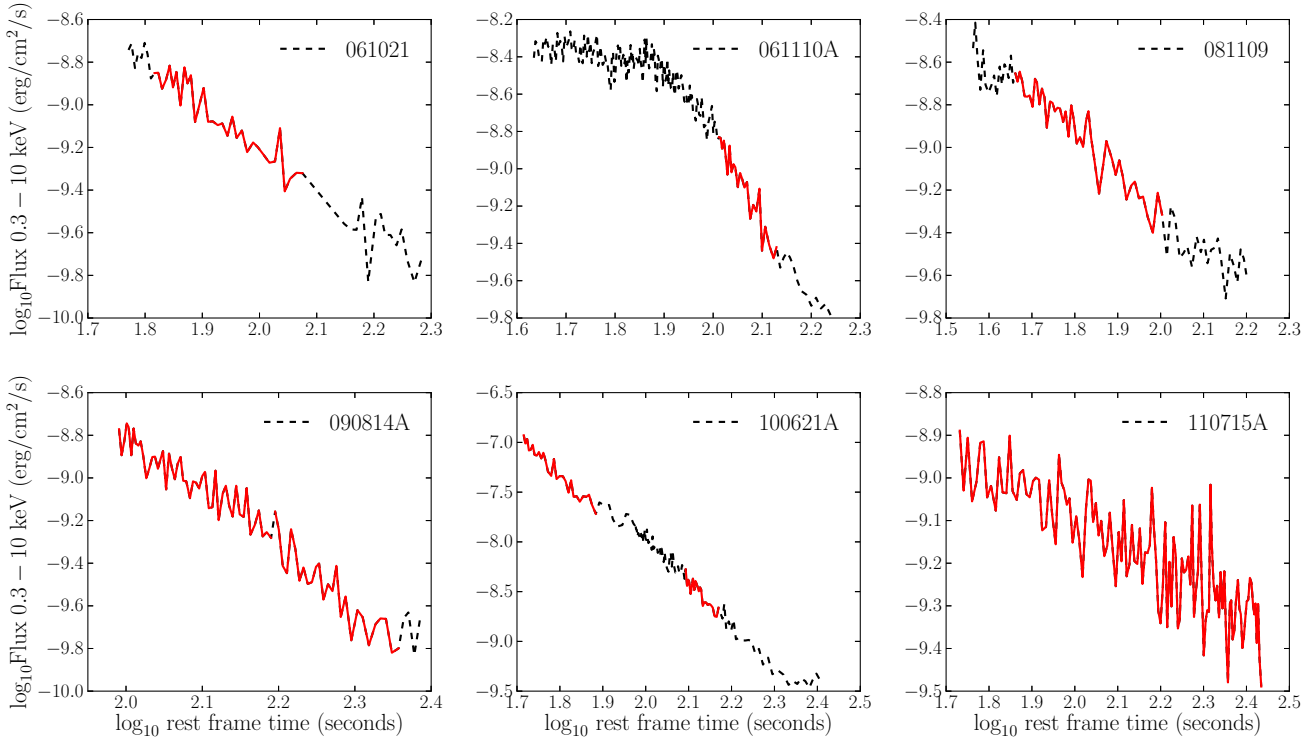}
\caption{The light curves for each of the bursts with probable excess emission. The curve is red (and solid) in the intervals selected for a time sliced analysis.}
\label{Fig_xrt_timeslices}
\end{figure*}

Figure~\ref{Fig_xrt_timeslices} shows the parts of the light curves which were used in the time sliced analysis of the final candidates. We tried to avoid achromatic and X-ray flares and effects from flattening of the light curves, when the time intervals were selected, but we note that evolution of the spectral parameters, the power law index and/or the black body temperature and normalisation, may be occurring during any of our spectra. When analysing time sliced spectra this risk is lowered, and we examined the hardness ratios in the XRT Repository in order to flag incidences of strong spectral evolution. GRBs 061021, 061110A, 090814A and 110715A have approximately constant hardness ratios in the intervals where time sliced spectra were analysed. GRB\,081109 shows a slowly hardening spectrum, while GRB\,100621A becomes significantly softer during the first time slice but shows no spectral evolution during our second time slice. Therefore the parameters we derive in Table~\ref{table:Radii} for these latter two GRBs should be considered more uncertain than the error bars allow. Evolution of the black body component has been demonstrated for GRBs 060218 \citep[e.g. ][]{2006Natur.442.1008C}, 090618 \citep{2011MNRAS.416.2078P}, 100316D \citep{2011MNRAS.411.2792S,2012A&A...539A..76O} and 101219B (Paper I), where the black body cools and expands with time, but typically this evolution is slow compared with the exposure times covered by our X-ray spectra.

\subsection{Contribution from the prompt emission}

It is important that the prompt emission, which has a spectrum peaking at the energy, $E_\text{peak}$, is not entering our X-ray spectrum for the \emph{Swift} XRT ($0.3-10$ keV), since this potentially could give features resembling black body components. Here is a summary of the measurement of $E_\text{peak}$ for each burst:

\paragraph*{061021} $E_\text{peak}$ was measured to lie above 540 keV up to 8 seconds after the trigger \citep{2006GCN..5748....1G}. This is well above XRT-energies.

\paragraph*{061110A} No spectral break or cutoff was seen with \emph{Swift}/BAT, so no $E_\text{peak}$ information is present.  The role of the prompt emission remains unknown for this burst.

\paragraph*{090814A} No spectral break or cutoff was seen by \emph{Swift}/BAT. The role of the prompt emission remains unknown for this burst.

\paragraph*{081109} Fermi/GBM measured $E_\text{peak}=240\pm 60$ keV \citep{2008GCN..8505....1V} after 26 seconds. Our spectrum is well below that energy.

\paragraph*{100621A} Konus-Wind fits their spectrum covering $0-74$ seconds with $E_\text{peak} = 95_{-13}^{+18}$ keV \citep{2010GCN..10882...1G}. Our spectrum starts at 80 s, so probably the peak energy was still well above the XRT band.

\paragraph*{110715A} 18 seconds after trigger Konus-wind measured $E_\text{peak}=120^{+12}_{-11}$ keV \citep{2011GCN..12166...1G}, and BAT found $120\pm 21$ keV \citep{2011GCN..12160...1U}. It is unlikely, that this affects our XRT spectrum, which starts at 98 seconds.

\begin{table*}

\label{table:Radii}
\centering 
\caption{Radii, temperatures and luminosities of the black body components of the final candidates. Candidates from this work and other works are shown. If a value is {\bf{bold}}, it has been calculated using Stefan-Boltzmann's law (see Eq. \eqref{StefanBoltzmann}), from the two non-bold values for a given candidate. If a $E_\text{iso}$-value is \underline{underlined}, it has been calculated from the fluence in the 15-150 keV band and it should only be seen as a lower limit.
}
\begin{tabularx}{\linewidth}{lrcrc|lcrr} 
\hline\hline 
GRB &Radius (m) & $kT$ (keV) &$L$ ($10^{47}$ erg s$^{-1}$) &ref&$z$& Time (s)&$T_{90}$ (s)&$E_\text{iso}$ (erg)\\
\hline
061021  & $\bf{6.6\times 10^{10}}$ & 0.13 &$1.4$&this work&$0.3663$&$87-170$&46&$4.6\times 10^{51}$ \\
061110A  & $\bf{1.4\times 10^{11}}$ & 0.19 &$25.6$& this work&$0.7578$&$180-240$&41& \underline{$2.9\times 10^{51}$} \\
081109  & $\bf{2.0\times 10^{11}}$&0.16&$27.1$& this work & $0.9787$&$90-200$&190& \underline{$1.8\times 10^{52}$}  \\
090814A   & $\bf{3.1\times 10^{10}}$ & 0.30 &$8.4$& this work&0.696&$166-265$&80&\underline{$2.8\times 10^{51}$}\\
090814A  & $\bf{4.7\times 10^{10}}$ & 0.20 &$3.7$&this work&0.696&$165-390$&80&\underline{$2.8\times 10^{51}$}\\
100621A &$\bf{8.7\times 10^{10}}$ & 0.25 &$32.4$&this work&0.542&$190-230$&64&$2.8 \times 10^{52}$\\
110715A & $\bf{2.7\times 10^{11}}$ & 0.15  &$ 43.0$ &this work&0.82&$97-501$&13&$4.1\times10^{52}$\\
\hline
060218 & $1.0\times 10^{10}$ & 0.20&$\bf{0.2}$&\citet{2006Natur.442.1008C}&$0.033$&200&2100&$6.2\times 10^{49}$\\
090618 & $3.0\times 10^{10}$&1.00&$1000.0$&\citet{2011MNRAS.416.2078P}&$0.54$&150  &113&$2.5\times 10^{53}$\\
090618  & $1.0\times 10^{11}$&0.20&$100.0$&\citet{2011MNRAS.416.2078P}&$0.54$&250 &113&$2.5\times 10^{53}$\\
100316D & $\bf{3\times 10^{10}}$ & $0.14$ &0.3&\citet{2011MNRAS.411.2792S}&0.0591&250&$>1300$&$3.9\times 10^{49}$\\
101219B & $\bf{2.4\times 10^{10}}$&0.20&$1.0$& Paper I &$0.5519$&250&34&$4.2 \times 10^{51}$\\
\hline\hline
\end{tabularx}

\end{table*}

\subsection{Is it a black body component?}

The six bursts in our final candidate list (GRB~061021, GRB~061110A, GRB~081109, GRB~090814A, GRB~100621A and GRB~110715A) are clearly special, since models with a black body component included give better fits than absorbed power laws. We do, however, not have conclusive evidence that the actual emission mechanism is a black body component. Other possible explanations could be a model with multiple absorption components at different redshifts, the prompt emission peak could be moving through the band, there could be strong spectral evolution, or the determined redshift could be wrong. We have a total of 190 bursts in our sample, so some of them will likely be affected by such features. Due to these caveats we will not report \emph{discovery of black body emission}, but instead we will report the finding of bursts with \emph{possible black body emission}.

\section{Properties of the black body components}

In this section, we will assume, that the actual mechanism behind the excess emission in the spectra of the six bursts is a single temperature black body component. Table~\ref{table:Radii} shows radius (assuming spherical symmetry) and temperature for these bursts together with values reported in other studies. For our final candidates the radii are calculated from the fitted temperature and luminosity. In the calculation we assume a spherically symmetric black body, in thermal equilibrium, which can be described by the Stefan-Boltzmann law,
\begin{align}
L(\text{erg } \text{s}^{-1})=1.105\times 10^{29}\times r^2(\text{m})\times T^4(\text{keV})\label{StefanBoltzmann}
\end{align}
Also note that the luminosities and radii in the table are lower limits, since only the photons in a $0.3-10$ keV are included in the luminosity calculation. Due to these caveats, the values in the table should be seen as rough estimates only.

The bursts have temperatures in the range, $0.1-0.3$ keV, which is consistent with the previously proposed black body components in the time-averaged WT spectra of GRB~060218, 090618, 100316D and 101219B. They also have luminosities consistent with these previously studied bursts, whereas the radii are sligthly larger for GRB 061110A, 081109 and 110715A than the previously claimed examples. 

In the case of GRB~060218 several studies \citep{2007MNRAS.382L..77G,2007MNRAS.375L..36G,2008ApJ...683L.135C,2007MNRAS.375..240L} disfavour the scenario that the excess emission is due to a black body component, since they can not explain the large black body luminosities. All our bursts with possible black body emission have a luminosity larger than GRB~060218 (Table~\ref{table:Radii}), so they might suffer from a similar problem.  

All the bursts have durations ($T_{90}$) between 13 and 190 seconds, which is typical for long bursts. The redshifts are in the range, $z=0.37-0.98$, so they are slightly more distant than the cases reported in other studies ($z=0.03-0.55$).

\subsection{The fraction of bursts with black body components}\label{fractionofbursts}

In the Model 1 to 2 comparison we found three new candidates (GRB~061021, 061110A, 090814A) and we re-discovered GRB~060218, 090618, 101219B. So we found a black body component in 6 out of the 116 bursts (i.e. 5\%) with successful fits.

\begin{figure}
\centering
\includegraphics[width = 0.45 \textwidth]{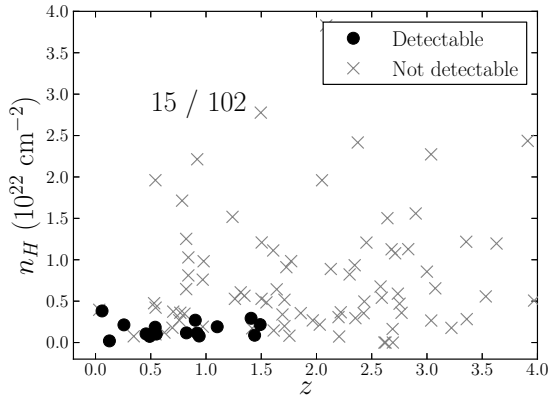}
\caption{$n_H$ versus $z$ for all the successfully fitted bursts. If a black body component, like the one present in GRB~101219B, is detectable in a GRB it is marked with a \emph{black circle}. In 15 out of 102 bursts it is possible to recover such a black body component.}
\label{Fig_nHz}
\end{figure}

We will now shed light on the fraction of \emph{Swift} bursts for which it is possible to recover a black body component like the one in GRB~101219B. In Section~\ref{Recovering} it was found that such a black body component can be detected (with the F-test comparison of Model 1 and 2) if $z<1.5$, $n_H < 0.4 \times 10^{22}$ cm$^{-2}$ and the logarithm of the powerlaw normalisation is below $-0.5$ (see Figure See Figure~\ref{Fig_101219B_nH}, \ref{Fig_Wrongfix_z_exptime} and \ref{Fig_101219B_Gamma}). Figure~\ref{Fig_nHz} shows $n_H$ versus $z$ for all the bursts in our sample with successful fits. It is marked whether a bursts passes these detection criteria. We see that in 15 $\%$, 15 out of 102 bursts, a detection will be possible.
This calculation is, of course, only a rough estimate. The X-ray black body components need not all be like the one in GRB 101219B, and the limits we have used on $z$, $n_H$, $\Gamma$ and power law normalisation are expected to be degenerate with each other. This calculation is, however, sufficient to establish that black body components only are detectable in a fraction of the bursts, even if such a black body component is present in all GRBs.

None of the new candidates found in this work have a optically confirmed SN. Note, however, that we excluded bursts which already are discussed in other works, including Paper I where several bursts with associated SNe are analysed.

\section{Conclusion}

We have first examined under which conditions a black body component, like the one proposed to be in GRB~101219B, can be recovered. At high redshift and in environments with intermediate or high column densities a detection of such a component will not be possible. We also find that it will be hard to recover a black body component, when a bright afterglow emission is present. We show that detection of a black body component only will be possible in a small fraction of all GRBs in our sample.

We have searched for black body components in all the \emph{Swift} bursts with known redshift and created a list of bursts with possible black body components (GRB~061021, 061110A, 081109, 090814A, 100621A and 110715A). They have temperatures, radii and luminosities similar to those found in previous studies of black body components in GRBs.

\section*{Acknowledgments}
We wish to thank J.P.U. Fynbo for commenting on the paper. This work was made possible through a University of Leicester College of Science Fellowship Support Scheme.
The Dark Cosmology Centre is funded by the Danish National Research Foundation. RLCS acknowledges support from a Royal Society Fellowship.

\def\aj{AJ}
\def\araa{ARA\&A}
\def\apj{ApJ}
\def\apjl{ApJ}
\def\apjs{ApJS}
\def\apss{Ap\&SS}
\def\aap{A\&A}
\def\aapr{A\&A~Rev.}
\def\aaps{A\&AS}
\def\mnras{MNRAS}
\def\nat{Nature}
\def\pasp{PASP}
\def\aplett{Astrophys.~Lett.}
\def\physrep{Physical Reviews}

\bibliographystyle{mn2e}
%\bibliography{ref}

\begin{thebibliography}{}

\bibitem[\protect\citeauthoryear{{Barthelmy} et~al.,}{{Barthelmy}
  et~al.}{2005}]{2005Natur.438..994B}
{Barthelmy} S.~D.,  et~al., 2005, \nat, 438, 994

\bibitem[\protect\citeauthoryear{{Burrows} et~al.,}{{Burrows}
  et~al.}{2005}]{2005SSRv..120..165B}
{Burrows} D.~N.,  et~al., 2005, SSRv, 120, 165

\bibitem[\protect\citeauthoryear{{Butler}}{{Butler}}{2007}]{2007ApJ...656.1001%
B}
{Butler} N.~R.,  2007, \apj, 656, 1001

\bibitem[\protect\citeauthoryear{{Campana} et~al.,}{{Campana}
  et~al.}{2006}]{2006Natur.442.1008C}
{Campana} S.,  et~al., 2006, \nat, 442, 1008

\bibitem[\protect\citeauthoryear{{Cano} et~al.,}{{Cano}
  et~al.}{2011}]{2011MNRAS.413..669C}
{Cano} Z.,  et~al., 2011, \mnras, 413, 669

\bibitem[\protect\citeauthoryear{{Cash}}{{Cash}}{1979}]{1979ApJ...228..939C}
{Cash} W.,  1979, \apj, 228, 939

\bibitem[\protect\citeauthoryear{{Chevalier} \& {Fransson}}{{Chevalier} \&
  {Fransson}}{2008}]{2008ApJ...683L.135C}
{Chevalier} R.~A.,  {Fransson} C.,  2008, \apjl, 683, L135

\bibitem[\protect\citeauthoryear{{Cobb}, {Bloom}, {Perley}, {Morgan}, {Cenko}
  \& {Filippenko}}{{Cobb} et~al.}{2010}]{2010ApJ...718L.150C}
{Cobb} B.~E.,  {Bloom} J.~S.,  {Perley} D.~A.,  {Morgan} A.~N.,  {Cenko} S.~B.,
     {Filippenko} A.~V.,  2010, \apjl, 718, L150

\bibitem[\protect\citeauthoryear{{Cucchiara} et~al.,}{{Cucchiara}
  et~al.}{2011}]{2011ApJ...736....7C}
{Cucchiara} A.,  et~al., 2011, \apj, 736, 7

\bibitem[\protect\citeauthoryear{{de Ugarte Postigo} et~al.,}{{de Ugarte
  Postigo}  et~al.}{2011}]{2011GCN..11579...1D}
{de Ugarte Postigo} A.,  et~al., 2011, GRB Coordinates Network, 11579, 1

\bibitem[\protect\citeauthoryear{{Evans} et~al.,}{{Evans}
  et~al.}{2009}]{2009MNRAS.397.1177E}
{Evans} P.~A.,  et~al., 2009, \mnras, 397, 1177

\bibitem[\protect\citeauthoryear{{Fynbo} et~al.,}{{Fynbo}
  et~al.}{2009}]{2009ApJS..185..526F}
{Fynbo} J.~P.~U.,  et~al., 2009, \apjs, 185, 526

\bibitem[\protect\citeauthoryear{{Galama} et~al.,}{{Galama}
  et~al.}{1998}]{1998Natur.395..670G}
{Galama} T.~J.,  et~al., 1998, \nat, 395, 670

\bibitem[\protect\citeauthoryear{{Ghisellini}, {Ghirlanda} \&
  {Tavecchio}}{{Ghisellini} et~al.}{2007a}]{2007MNRAS.382L..77G}
{Ghisellini} G.,  {Ghirlanda} G.,    {Tavecchio} F.,  2007a, \mnras, 382, L77

\bibitem[\protect\citeauthoryear{{Ghisellini}, {Ghirlanda} \&
  {Tavecchio}}{{Ghisellini} et~al.}{2007b}]{2007MNRAS.375L..36G}
{Ghisellini} G.,  {Ghirlanda} G.,    {Tavecchio} F.,  2007b, \mnras, 375, L36

\bibitem[\protect\citeauthoryear{{Golenetskii}, {Aptekar}, {Mazets},
  {Pal'Shin}, {Frederiks} \& {Cline}}{{Golenetskii}
  et~al.}{2006}]{2006GCN..5748....1G}
{Golenetskii} S.,  {Aptekar} R.,  {Mazets} E.,  {Pal'Shin} V.,  {Frederiks} D.,
     {Cline} T.,  2006, GRB Coordinates Network, 5748, 1

\bibitem[\protect\citeauthoryear{{Golenetskii} et~al.,}{{Golenetskii}
  et~al.}{2010}]{2010GCN..10882...1G}
{Golenetskii} S.,  et~al., 2010, GRB Coordinates Network, 10882, 1

\bibitem[\protect\citeauthoryear{{Golenetskii} et~al.,}{{Golenetskii}
  et~al.}{2011}]{2011GCN..12166...1G}
{Golenetskii} S.,  et~al., 2011, GRB Coordinates Network, 12166, 1

\bibitem[\protect\citeauthoryear{{Hjorth} \& {Bloom}}{{Hjorth} \&
  {Bloom}}{2011}]{2011arXiv1104.2274H}
{Hjorth} J.,  {Bloom} J.~S.,  2011, ArXiv e-prints astroph/1104.2274

\bibitem[\protect\citeauthoryear{{Hjorth} et~al.,}{{Hjorth}
  et~al.}{2003}]{2003Natur.423..847H}
{Hjorth} J.,  et~al., 2003, \nat, 423, 847

\bibitem[\protect\citeauthoryear{{Jakobsson} et~al.,}{{Jakobsson}
  et~al.}{2009}]{2009GCN..9797....1J}
{Jakobsson} P.,  et~al., 2009, GRB Coordinates Network, 9797, 1

\bibitem[\protect\citeauthoryear{{Kalberla}, {Burton}, {Hartmann}, {Arnal},
  {Bajaja}, {Morras} \& {P{\"o}ppel}}{{Kalberla}
  et~al.}{2005}]{2005A&A...440..775K}
{Kalberla} P.~M.~W.,  {Burton} W.~B.,  {Hartmann} D.,  {Arnal} E.~M.,  {Bajaja}
  E.,  {Morras} R.,    {P{\"o}ppel} W.~G.~L.,  2005, \aap, 440, 775

\bibitem[\protect\citeauthoryear{{Kouveliotou}, {Meegan}, {Fishman}, {Bhat},
  {Briggs}, {Koshut}, {Paciesas} \& {Pendleton}}{{Kouveliotou}
  et~al.}{1993}]{1993ApJ...413L.101K}
{Kouveliotou} C.,  {Meegan} C.~A.,  {Fishman} G.~J.,  {Bhat} N.~P.,  {Briggs}
  M.~S.,  {Koshut} T.~M.,  {Paciesas} W.~S.,    {Pendleton} G.~N.,  1993,
  \apjl, 413, L101

\bibitem[\protect\citeauthoryear{{Kr{\"u}hler} et~al.,}{{Kr{\"u}hler}
  et~al.}{2011a}]{2011A&A...526A.153K}
{Kr{\"u}hler} T.,  et~al., 2011a, \aap, 526, A153

\bibitem[\protect\citeauthoryear{{Kr{\"u}hler} et~al.,}{{Kr{\"u}hler}
  et~al.}{2011b}]{2011AA...534A.108K}
{Kr{\"u}hler} T.,  et~al., 2011b, \aap, 534, A108

\bibitem[\protect\citeauthoryear{{Lampton}, {Margon} \& {Bowyer}}{{Lampton}
  et~al.}{1976}]{1976ApJ...208..177L}
{Lampton} M.,  {Margon} B.,    {Bowyer} S.,  1976, \apj, 208, 177

\bibitem[\protect\citeauthoryear{{Li}}{{Li}}{2007}]{2007MNRAS.375..240L}
{Li} L.-X.,  2007, \mnras, 375, 240

\bibitem[\protect\citeauthoryear{{Lipkin} et~al.,}{{Lipkin}
  et~al.}{2004}]{2004ApJ...606..381L}
{Lipkin} Y.~M.,  et~al., 2004, \apj, 606, 381

\bibitem[\protect\citeauthoryear{{Mazzali}, {Deng}, {Nomoto}, {Sauer}, {Pian},
  {Tominaga}, {Tanaka}, {Maeda} \& {Filippenko}}{{Mazzali}
  et~al.}{2006}]{2006Natur.442.1018M}
{Mazzali} P.~A.,  {Deng} J.,  {Nomoto} K.,  {Sauer} D.~N.,  {Pian} E.,
  {Tominaga} N.,  {Tanaka} M.,  {Maeda} K.,    {Filippenko} A.~V.,  2006, \nat,
  442, 1018

\bibitem[\protect\citeauthoryear{{Milvang-Jensen} et~al.,}{{Milvang-Jensen}
  et~al.}{2010}]{2010GCN..10876...1M}
{Milvang-Jensen} B.,  et~al., 2010, GRB Coordinates Network, 10876, 1

\bibitem[\protect\citeauthoryear{{Olivares E.} et~al.,}{{Olivares E.}
  et~al.}{2012}]{2012A&A...539A..76O}
{Olivares E.} F.,  et~al., 2012, \aap, 539, A76

\bibitem[\protect\citeauthoryear{{Page}, {Starling}, {Fitzpatrick}, {Pandey},
  {Osborne} et~al.,}{{Page} et~al.}{2011}]{2011MNRAS.416.2078P}
{Page} K.~L.,  {Starling} R.~L.~C.,  {Fitzpatrick} G.,  {Pandey} S.~B.,
  {Osborne} J.~P.,    et~al., 2011, \mnras, 416, 2078

\bibitem[\protect\citeauthoryear{{Patat} et~al.,}{{Patat}
  et~al.}{2001}]{2001ApJ...555..900P}
{Patat} F.,  et~al., 2001, \apj, 555, 900

\bibitem[\protect\citeauthoryear{{Pian} et~al.,}{{Pian}
  et~al.}{2006}]{2006Natur.442.1011P}
{Pian} E.,  et~al., 2006, \nat, 442, 1011

\bibitem[\protect\citeauthoryear{{Piranomonte}, {Vergani}, {Malesani}, {Fynbo},
  {Wiersema} \& {Kaper}}{{Piranomonte} et~al.}{2011}]{2011GCN..12164...1P}
{Piranomonte} S.,  {Vergani} S.~D.,  {Malesani} D.,  {Fynbo} J.~P.~U.,
  {Wiersema} K.,    {Kaper} L.,  2011, GRB Coordinates Network, 12164, 1

\bibitem[\protect\citeauthoryear{{Protassov}, {van Dyk}, {Connors}, {Kashyap}
  \& {Siemiginowska}}{{Protassov} et~al.}{2002}]{2002ApJ...571..545P}
{Protassov} R.,  {van Dyk} D.~A.,  {Connors} A.,  {Kashyap} V.~L.,
  {Siemiginowska} A.,  2002, \apj, 571, 545

\bibitem[\protect\citeauthoryear{{Salvaterra} et~al.,}{{Salvaterra}
  et~al.}{2009}]{2009Natur.461.1258S}
{Salvaterra} R.,  et~al., 2009, \nat, 461, 1258

\bibitem[\protect\citeauthoryear{{Soderberg} et~al.,}{{Soderberg}
  et~al.}{2006}]{2006Natur.442.1014S}
{Soderberg} A.~M.,  et~al., 2006, \nat, 442, 1014

\bibitem[\protect\citeauthoryear{{Sparre} et~al.,}{{Sparre}
  et~al.}{2011}]{2011ApJ...735L..24S}
{Sparre} M.,  et~al., 2011, \apjl, 735, L24

\bibitem[\protect\citeauthoryear{{Starling} et~al.,}{{Starling}
  et~al.}{2011}]{2011MNRAS.411.2792S}
{Starling} R.~L.~C.,  et~al., 2011, \mnras, 411, 2792

\bibitem[\protect\citeauthoryear{{Starling} et~al.,}{{Starling}
  et~al.}{2012}]{Starling2012}
{Starling} R.~L.~C.,  et~al., 2012, ArXiv e-prints, astroph/1207.1444

\bibitem[\protect\citeauthoryear{{Starling}, {Page} \& {M.~Sparre}}{{Starling}
  et~al.}{2012}]{IAU}
{Starling} R.~L.~C.,  {Page} K.~L.,    {M.~Sparre} 2012, Accepted to the
  proceedings of IAU 279 'Death of massive stars: supernovae and gamma-ray
  bursts'

\bibitem[\protect\citeauthoryear{{Tanvir} et~al.,}{{Tanvir}
  et~al.}{2009}]{2009Natur.461.1254T}
{Tanvir} N.~R.,  et~al., 2009, \nat, 461, 1254

\bibitem[\protect\citeauthoryear{{Ukwatta} et~al.,}{{Ukwatta}
  et~al.}{2011}]{2011GCN..12160...1U}
{Ukwatta} T.~N.,  et~al., 2011, GRB Coordinates Network, 12160, 1

\bibitem[\protect\citeauthoryear{{Verner}, {Ferland}, {Korista} \&
  {Yakovlev}}{{Verner} et~al.}{1996}]{1996ApJ...465..487V}
{Verner} D.~A.,  {Ferland} G.~J.,  {Korista} K.~T.,    {Yakovlev} D.~G.,  1996,
  \apj, 465, 487

\bibitem[\protect\citeauthoryear{{von Kienlin}}{{von
  Kienlin}}{2008}]{2008GCN..8505....1V}
{von Kienlin} A.,  2008, GRB Coordinates Network, 8505, 1

\bibitem[\protect\citeauthoryear{{Waxman}, {M{\'e}sz{\'a}ros} \&
  {Campana}}{{Waxman} et~al.}{2007}]{2007ApJ...667..351W}
{Waxman} E.,  {M{\'e}sz{\'a}ros} P.,    {Campana} S.,  2007, \apj, 667, 351

\bibitem[\protect\citeauthoryear{{Wilms}, {Allen} \& {McCray}}{{Wilms}
  et~al.}{2000}]{2000ApJ...542..914W}
{Wilms} J.,  {Allen} A.,    {McCray} R.,  2000, \apj, 542, 914

\end{thebibliography}

\bsp

.
\label{lastpage}

\end{document}